\documentclass{mem}
\usepackage{natbib}\usepackage{txfonts}
\usepackage{graphicx}
\usepackage[a4paper]{hyperref}
\begin{document}
\def\teff{$T\rm_{eff }$}
\def\kms{$\mathrm {km s}^{-1}$}
\def\fun#1#2{\lower3.6pt\vbox{\baselineskip0pt\lineskip.9pt
\ialign{$\mathsurround=0pt#1\hfil##\hfil$\crcr#2\crcr\sim\crcr}}}
\def\lap{\mathrel{\mathpalette\fun <}}
\def\gap{\mathrel{\mathpalette\fun >}}
\def\beq{\begin{equation}}
\def\eeq{\end{equation}}

\title{
Black Holes and Nuclear Dynamics
}

   \subtitle{}

\author{
D. \,Merritt\inst{1} 
          }

  \offprints{D. Merritt}

\institute{
Department of Physics, 
Rochester Institute of Technology,
Rochester, NY 14623, USA
\email{merritt@astro.rit.edu}
}

\authorrunning{Merritt }

\titlerunning{Black Hole Nuclei}

\abstract{
Supermassive black holes inhabit galactic nuclei,
and their presence influences in crucial ways the
evolution of the stellar distribution.
The low-density cores observed in bright galaxies are
probably a result of black hole infall,
while steep density cusps like those at the
Galactic center are a result of energy
exchange between stars moving in the
gravitational field of the single black hole.
Loss-cone dynamics are substantially more
complex in galactic nuclei than in collisionally-relaxed
systems like globular clusters
due to the wider variety of possible geometries
and orbital populations.
The rate of star-black hole interactions has
begun to be constrained through observations
of energetic events associated with stellar tidal
disruptions.
}
\maketitle{}

\section{Introduction}

The association of supermassive black holes (SBHs) 
with galactic nuclei began even before their
observational confirmation, when \citet{novikov-64}
and \citet{salpeter-64} first discussed the growth
of massive objects at the centers of galaxies.
Since then, the mutual interaction of SBHs and stars
has played a central role both in the detection
and mass determination of SBHs \citep{ff-05}, 
and also in our theoretical understanding of how 
nuclei form and evolve.
The realization that galactic spheroids grow through
mergers added an interesting complication to this
picture, since galaxy mergers imply the formation
of binary SBHs \citep{BBR-80}, which can influence the stellar
distribution on larger spatial scales than single SBHs,
and which may be directly or indirectly
connected with nuclear activity \citep{sk-03}.

\section{Binary Black Holes and Cores}

Galaxy mergers bring SBHs together, and 
unless the binary SBHs themselves coalesce,
ejections will occur when a third SBH is deposited 
into a nucleus containing an uncoalesced binary 
\citep{MV-90,VHM-03}.
Such events can not be too frequent or the tight,
empirical correlations between SBH masses and
galaxy properties \citep{FM-00,graham-01,MH-03}
would be violated \citep{HK-02}.

\begin{figure}[]
\resizebox{\hsize}{!}{\includegraphics[clip=true]{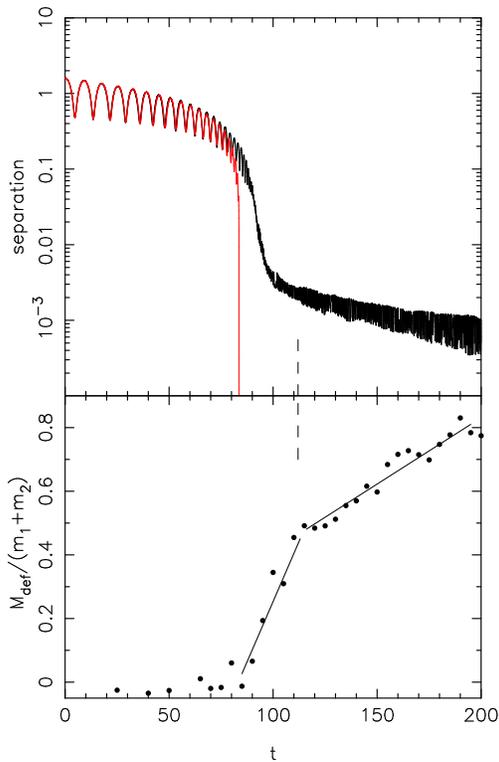}}
\caption{
\footnotesize
Short-term evolution of binary black holes in spherical galaxies.
{\it Upper panel:} Thick (black) line shows infall of 
a hole of mass $m_2$ into a nucleus containing
a hole of mass $m_1=10m_2$.
Thin (red) line is the orbital decay predicted by the
dynamical friction equation, assuming an unchanging galaxy.
The vertical dashed line indicates the time when $a\approx a_h$.
{\it Lower panel:} Evolution of the mass deficit, 
i.e. the mass displaced by the binary,  in units of the
combined mass of the two black holes.
The evolution slows at $a\approx a_h$ due to ejection
of stars on orbits that intersect the binary,
and thereafter the evolution rate is strongly
dependent on the number $N$ of particles used to
represent the galaxy (here, $N=2\times 10^5$).
For much larger $N$, the binary would stall at $a\approx a_h$
(see Fig. 2).
}
\label{fig:decay}
\end{figure}

Ultimately, coalescence is driven by emission
of gravitational waves, but this process only
becomes efficient when separations fall below
$\sim 10^{-3}$ pc.
By comparison, the natural separation between
two SBHs of mass $m_1$ and $m_2$ in a galactic nucleus is
\beq
a_h \equiv {G\mu\over 4\sigma^2}
\approx 2.7\ {\rm pc} {q\over (1+q)^2} 
M_{\bullet,8} \sigma_{200}^{-2}
\label{eq:ah}
\eeq
where $M_\bullet=m_1+m_2$,
$\mu= m_1m_2/M_\bullet$ is the reduced mass,
$q=m_2/m_1\le 1$ is the mass ratio, 
$M_{\bullet,8}=M_\bullet/10^8M_\odot$,
and $\sigma_{200}$ is the nuclear velocity dispersion
in units of $200$ km s$^{-1}$.
A binary with semi-major axis $a\approx a_h$ is ``hard,''
i.e. its binding energy per unit mass is $\sim\sigma^2$
and its total binding energy $E_h$ is a fraction
$\sim M_\bullet/M_{gal}\approx 10^{-3}$ that of the 
host galaxy.

\begin{figure}[t]
\resizebox{\hsize}{!}{\includegraphics[clip=true]{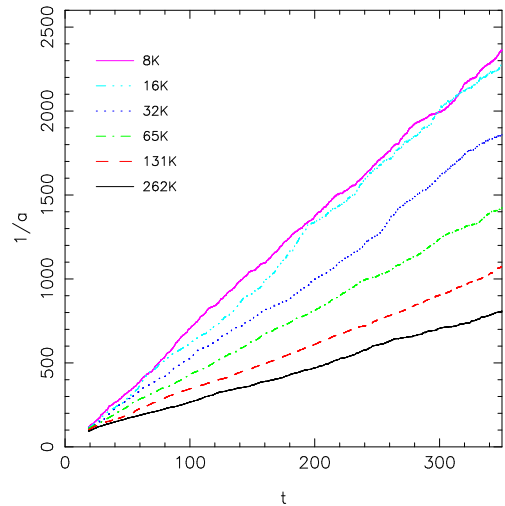}}
\caption{
\footnotesize
Long-term evolution of binary black holes in spherical galaxies.
Each curve shows the evolution of the inverse semi-major axis 
of an equal-mass binary in a spherical galaxy with an initially
power-law nuclear density profile.
Curves were obtained by averaging multiple 
$N$-body integrations with the same $N$.
As $N$ increases, the binary's hardening rate drops,
implying very slow evolution past $a\approx a_h$ 
in a real galaxy 
(adapted from Merritt, Mikkola \& Szell 2006).
}
\label{fig:lt}
\end{figure}

The separation $a_h$ is natural one since a binary with
$a\lap a_h$ efficiently ejects stars on intersecting orbits
\citep{MV-92,quinlan-96}.
In a spherical or axisymmetric galaxy, the total mass associated
with stars on such orbits is small, $\lap M_\bullet$, and
once the binary has ejected these stars, 
its evolution greatly slows.
This is the ``final parsec problem.''
In $N$-body simulations
(e.g. Milosavljevic \& Merritt 2001, Makino \& Funato 2004), 
the binary often {\it does}
evolve past $a\approx a_h$ but at a rate that is
strongly $N$-dependent (Figs.~\ref{fig:decay},\ref{fig:lt}).
This reflects the fact that processes 
like gravitational scattering and Brownian motion
that can refill a binary's loss cone are $m_\star$-dependent.

\begin{figure}[t!]
\resizebox{\hsize}{!}{\includegraphics[clip=true]{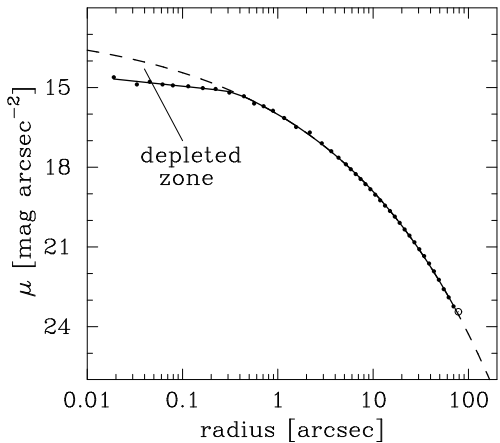}}
\caption{\footnotesize
  Observational determination of mass deficits.
 Plot shows the observed surface brightness profile of NGC 3348.
  The dashed line is a Sersic model fit to the large-radius
  data.  Solid line is the fit of an alternative model,
  the ``core-Sersic'' model, which fits both the inner and outer
  data well. The mass deficit is illustrated by the area
  designated ``depleted zone'' and the corresponding mass is
  roughly $3\times 10^8M_\odot$
(from Graham 2004).
}
\label{fig:sersic}
\end{figure}

Since the efficiency of loss-cone repopulation in
real galaxies is uncertain and may be low,
it is interesting to ask what the observable consequences
would  be of binaries stalling at $a\approx a_h$.
Using  the $M_\bullet-\sigma$ relation,
Eq.~(\ref{eq:ah}) can be written
\beq
a_{h,\rm pc}\approx 0.3\ q_{0.1}\ \sigma_{200}^{2.9}
\eeq
with $q_{0.1}\equiv q/0.1$.
``Dual'' SBHs are observed in a handful of interacting galaxies
and binary quasars \citep{mortlock-99,komossa-03,ballo-04},
but always with much larger separations, $\gap 1$ kpc.
True binaries with $a\approx a_h$ would be difficult to resolve
outside the Local Group even if both SBHs were luminous.
A number of active galaxies show evidence for variability
with periods similar to binary orbital periods,
but the variability may be a scaled-up version of 
the quasi-periodic oscillations associated with stellar-mass 
black holes and not the signature of a binary SBH \citep{uttley-05}.

\begin{figure*}[t]
\includegraphics[width=1.15\hsize]{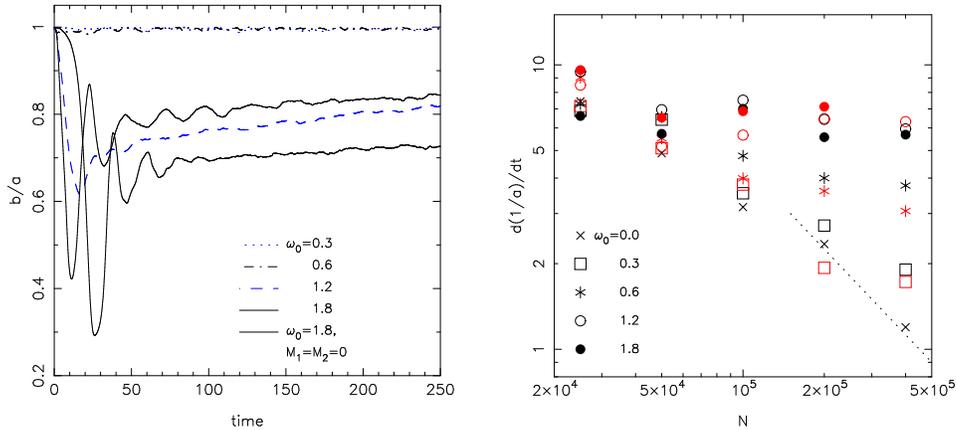}
\caption{
\footnotesize
Efficient merger of binary SBHs in barred galaxies.
The left panel shows the evolution of the bar amplitude
in a set of initially axisymmetric galaxy models with different
degreess of rotational support (measured by the parameter
$\omega_0$).
For $\omega_0\ge 0.6$ the models are bar-unstable.
The right panel shows the hardening rate of a massive binary 
at the center of the galaxy.
In the barred (triaxial) models, the binary's hardening
rate is independent of particle number $N$, while
in the axisymmetric models it falls roughly as 
$N^{-1}$ indicating a collisionally-resupplied loss cone
(adapted from Berczik et al. 2006)
}
\label{fig:triax}
\end{figure*}

An indirect way to track the evolution of a binary SBH
is via its effects on the structure of the nucleus.
The infalling hole begins displacing stars when it
first becomes bound to the second, at a separation
$\sim r_h\equiv GM_\bullet/\sigma^2 \approx 
10M_{\bullet,8}\sigma_{200}^{-2}\ {\rm pc} 
\approx 15M_{\bullet,8}^{0.59}\ {\rm pc}$.
Evolution from $r_h$ to $a_h$ is rapid unless the mass
ratio is extreme, and the ``damage'' done by the binary
to the nucleus can be robustly  estimated from $N$-body
simulations, independent of uncertainties about
the efficiency of loss-cone refilling (Fig.~\ref{fig:decay}).
A standard definition of the damage is the
``mass deficit'' $M_{def}$, the difference in 
integrated mass between the observed density profile
and the primordial (pre-merger) profile
\citep{mm-02}.
$N$-body simulations like those in Fig. 1
show that $M_{def}(a=a_h)\approx 0.5(m_1+m_2)$,
decreasing only weakly with $m_2$,
and nearly independent of the form of the initial
density profile.
This is a consequence of the scaling of $E_h$ 
with $m_1+m_2$ discussed above.
An interesting corollary is that the mass deficit
after ${\cal N}$ mergers should be 
$\sim 0.5 M_\bullet {\cal N}$,
with $M_\bullet$ the final (i.e. current) 
mass of the SBH, almost independent of
the details of the merger tree.

Bright ($M_B\lap -20.5$) elliptical galaxies
always exhibit cores, regions near
the center where the luminosity profile is nearly flat
(Ferrarese et al. 2006 and references therein).
Estimating $M_{def}$ in a real galaxy is not
straightforward however, since one does not know what
the pre-existing stellar distribution was.
A conservative approach is to
fit a smooth function, e.g. S\'ersic's law,
to the outer luminosity profile and extend it
inward; the mass deficit is then defined in 
terms of the differential profile (Fig.~\ref{fig:sersic}).
Mass deficits derived in this way are found to 
be roughly equal to $M_\bullet$
\citep{graham-04,acs-6}
This is consistent with the relation given above
if ${\cal N}\approx 2$, in reasonable accord with
hierarchical galaxy formation models.
Alternatively, if the binary evolves all the way
to coalescence and if gravitational waves are emitted anisotropically
during the final plunge (the ``rocket'' effect; 
Favata et al. 2004),
the displaced hole will transfer energy to the nucleus
before falling back to the center, increasing the
mass deficit by some appreciable fraction of $M_\bullet$
\citep{merritt-04,copycat-04}.

Evolution past $a\approx a_h$ may or may not
leave an additional mark on a nucleus.
In small dense galaxies like M32, two-body (star-star)
relaxation can repopulate a binary's loss cone
at interesting rates, possibly driving the binary
to coalescence in $\lap 10$ Gyr \citep{yu-02,MM-03}.
$N$-body simulations like those in Fig.~\ref{fig:lt} 
show that the mass deficit
increases as $M_{def}\propto\ln(a_h/a)$ in such
cases, so that the observed value of $M_{def}$ does
not contain much information about the final value of $a$.
The bright elliptical galaxies that are observed to 
contain cores have relaxation times that are much
too long for collisional
loss cone repopulation to occur however.
A more efficient way to drive a binary to
coalescence is to imbed it in a strongly non-axisymmetric
(barred or triaxial) galaxy \citep{poon-04b,holley-06}.
Triaxial potentials can support large populations
of ``centrophilic'' orbits, allowing a binary
to harden well past $a_h$ by interacting with stars
even in the absence of collisional loss-cone refilling 
(Fig.~\ref{fig:triax}).
Most of the stars supplying mass to the binary under
these circumstances have orbital apocenters $\gg r_h$
and their depletion has almost no effect on the 
nuclear density profile.

A binary with $a\ll a_h$ ejects stars with velocities
large enough to expel them from the galaxy.
A handful of hypervelocity stars (HVSs) have been found
in the Galactic halo that are candidates for ejection
\citep{brown-05,hirsch-05,brown-06}.
These stars must have left the nucleus 
$\lap 10^8$ yr ago however, implying that
a binary SBH was present until quite recently.
Other models can produce HVSs in the absence of a second
SBH (e.g. Hills 1988; Gualandris et al. 2005).
Ejection by binary SBHs might also be responsible for
populations like the intergalactic planetary nebulae 
observed in the Virgo cluster \citep{holley-05}.
However, such models require that
the binary SBHs ejecting the nebulae
harden well beyond $a=a_h$, and also that 
the hardening is driven by star-binary 
interactions; both assumptions are open to question,
particularly in the case of the brightest Virgo
galaxies.

The nuclei of merging galaxies are expected to contain
some of the largest concentrations of dense gas in the
universe.
Binary-gas simulations fall into two categories
depending on whether the gas is modelled as ``hot,''
i.e. having a specific kinetic energy comparable
with that of the stars, or ``cold,''
e.g. in a thin disk.
The first case is much easier to treat;
if the mass in gas within $\sim a_h$ is comparable to $M_\bullet$,
the additional component of dynamical friction drag 
can bring the two SBHs together in much less than $10^9$ yr
\citep{escala-04,escala-05}.
However it is not clear that such massive, dense accumulations of
hot gas can be sustained.
In the case of binary-disk interactions
the difficulties are computational, e.g.
resolving the gas on scales smaller than the
radial wavelength of the density waves induced by the binary.
Much progress in this area is to be expected in coming
years.

\section{Nuclear Equilibria}

\begin{figure}[t!]
\resizebox{\hsize}{!}{\includegraphics[clip=true]{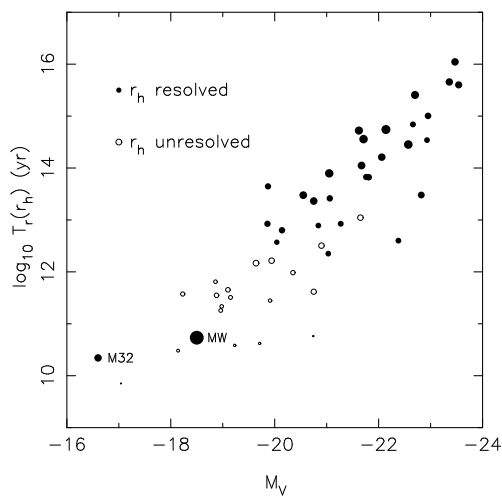}}
\caption{\footnotesize
Nuclear relaxation times in a sample of early-type
galaxies and bulges.
The size of the symbols reflects the degree to which
the SBH's influence radius was resolved;
filled symbols have $\theta_{r_h}>\theta_{obs}$ 
(resolved) and open circles have
$\theta_{r_h}<\theta_{obs}$ (unresolved)
(from Merritt \& Szell 2005).
}
\label{fig:tr}
\end{figure}

Once the massive binary has coalesced 
(or otherwise stopped interacting with stars),
the stellar distribution in the nucleus can
evolve to a steady state.
An important distinction can be made between
{\it collisional} nuclei, which have
two-body (star-star) relaxation times less than
$10$ Gyr, and {\it collisionless} nuclei,
with $T_r\gap 10$ Gyr.
Fig.~\ref{fig:tr} shows that $T_r(r_h)$
drops below $10$ Gyr in spheroids roughly as faint
as the bulge of the Milky Way.

The morphology of collisionless nuclei is constrained
only by Jean's theorem, i.e. by the requirement that the
density be representable as a superposition
of orbits integrated in the self-consistent potential.
Collisionless equilibria can take many forms,
including triaxial nuclei with chaotic orbits
\citep{poon-04a},
as well as the simpler, axisymmetric models that
are the basis for most SBH mass estimation
\citep{gebhardt-03}.
A recent review is given by Merritt (2006).

In a collisional nucleus, on the other hand,
exchange of energy between stars drives the phase-space
density toward a characteristic form.
Here we discuss the more limited class of solutions
associated with collisionally-relaxed nuclei.

Gravitational encounters drive the velocity distribution 
of stars around a black hole toward a Maxwellian on a time scale of $\sim T_r$,
but a Maxwellian velocity distribution implies
an exponentially divergent mass near the hole.
The existence of a region close to the black hole in which
stars are captured or destroyed prevents the nucleus
from reaching precise thermal equilibrium.
The density must drop to zero on orbits that intersect
the hole's event horizon or that pass within
the tidal disruption sphere at $r=r_t$; the latter
radius is most relevant here since galaxies
with collisional nuclei probably always have
$M_\bullet<10^8M_\odot$ (Fig.~\ref{fig:tr})
and hence $r_t>2GM_\bullet/c^2$.

For a single-mass population of stars moving
in the Keplerian potential of a black hole,
the steady-state solution to the isotropic,
orbit-averaged Fokker-Planck equation in the 
presence of a central ``sink'' is approximately
\begin{eqnarray}
f(E) &=& f_0|E|^{1/4},\ \ \rho(r) = \rho_0r^{-7/4}, \\ 
|E_h| &\ll& |E| \ll |E_t|, \ \ r_t\ll r \ll r_h
\end{eqnarray}
\citep{bw-76,LS-77}.
Here $E_t\equiv -GM_\bullet/r_t$ and $E_h\equiv -GM_\bullet/r_h$. 
This is a ``zero-flux'' solution, i.e. it implies
$F_E=0$ at $|E_h| \ll |E| \ll |E_t|$ where $F_E$
is the encounter-driven flux of stars in energy space.
The actual flux is small but non-zero, of order
\beq
F_E \approx {n(r_t)r_t^3\over T_r(r_t)} \propto r_t.
\label{eq:fluxe}
\eeq
In other words, the flux is limited by the rate
at which stars can diffuse into the disruption sphere at $r_t$.
This flux is ``small'' in the sense that the one-way
flux of stars in or out through a surface
at $r_t\ll r \ll r_h$ is much greater;
except near $r_t$, the inward and outward fluxes almost cancel.

The Bahcall-Wolf solution has been verified in a number of
studies based on Fokker-Planck \citep{CK-78},
fluid \citep{amaro-04}, or Monte-Carlo 
\citep{marchant-80,duncan-83,freitag-02}
approximations.
Most recently, advances in computer hardware
\citep{namura-03}
have made it possible to test the Bahcall-Wolf solution
via direct $N$-body integrations, 
avoiding the approximations
of the Fokker-Planck formalism 
\citep{preto-04,baumgardt-04,MS-05}.
Another advantage of this approach is that it can easily
deal with complex initial conditions and nonspherical geometries,
such as those set up by a binary SBH.

\begin{figure*}[t!]
\resizebox{\hsize}{!}{\includegraphics[clip=true,angle=-90.]{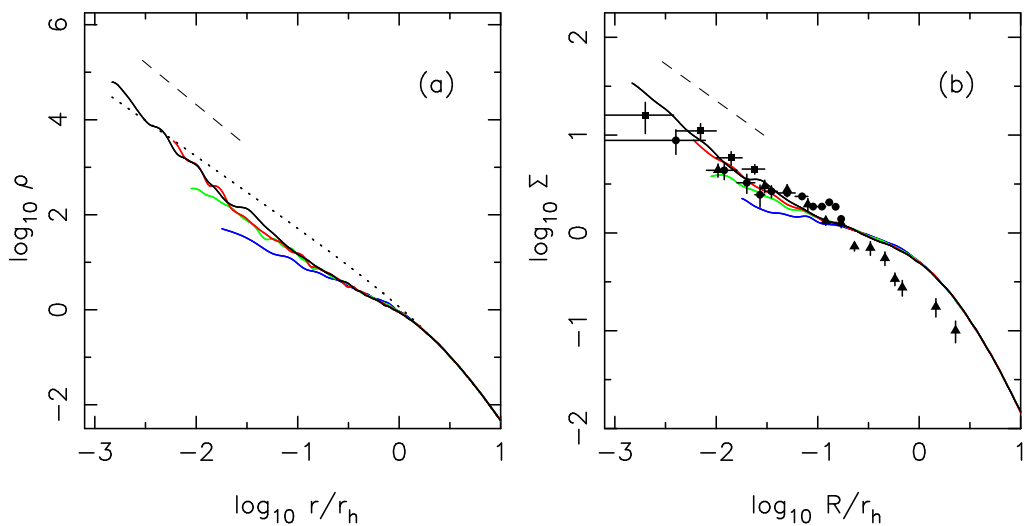}}
\caption{
\footnotesize
Cusp regeneration.
Curves show the evolution of the spatial (a) and projected
(b) density profiles of an $N$-body model in which
an initial, $\rho\sim r^{-1.5}$ density cusp (dotted line)
was destroyed by infall of a second black hole of
mass $m_2=0.5m_1$.
Blue (lower) line shows the density after infall,
and green, red and black lines show the evolving density
after the two black holes have been merged into one;
the final time is $\sim 10$ Gyr if scaled to the Milky
Way.
Symbols in (b) show the observed surface density of stars
near the Galactic Center (Genzel et al. 2003).
Dashed lines have logarithmic slopes of $-1.75$ (a)
and $-0.75$ (b), the Bahcall-Wolf ``zero flux'' solution
(from Merritt \& Szell 2005).
}
\label{fig:regen}
\end{figure*}

Fig.~\ref{fig:regen} shows a simulation of
cusp destruction (by a binary SBH) followed by
cusp regeneration (via the Bahcall-Wolf mechanism).
The Bahcall-Wolf cusp 
rises above the core inside a radius $\sim 0.2r_h$ and 
reaches its steady-state form in a time 
$2-3 T_r(0.2 r_h)$.
In the Milky Way nucleus, $0.2r_h\approx 0.7$ pc
and $T_r(0.2r_h)\approx 6$ Gyr 
(assuming $m_\star=0.7M_\odot$).
In fact, as Fig.~\ref{fig:regen} shows,
the stellar distribution at the Galactic
center is close to the Bahcall-Wolf form.

Figs.~\ref{fig:tr} and \ref{fig:regen} suggest that Bahcall-Wolf cusps
should be present in many galaxies with luminosities
similar to that of the Milky Way bulge, or fainter,
assuming that they contain SBHs.
M32 fits these criteria; the mass of its SBH
is uncertain by a factor $\sim 5$ \citep{valluri-04}
but even if the largest possible value is assumed,
$0.2r_h\approx 0.8\ {\rm pc} \approx 0.3''$
would only barely be resolved. 
NGC 205 clearly contains a constant-density 
core but also shows no dynamical evidence of a 
massive black hole \citep{valluri-05}.
All other galaxies in which $r_h$ is resolved
have nuclear relaxation times $\gg 10$ Gyr (Fig.~\ref{fig:tr}).

\section{Loss-Cone Dynamics}

As discussed above, the existence of a region $r\le r_t$
close to the SBH where stars are destroyed has a profound
influence on the steady-state distribution of stars 
even at $r\gg r_t$.
Loss of stars is also important because of its 
observational consequences: 
tidally disrupted stars are expected to produce
X- and UV radiation with luminosities of 
$\sim 10^{44}$ erg s$^{-1}$
\citep{silk-77,rees-90}.

Once a central SBH has removed stars on 
orbits that intersect the disruption sphere $r=r_t$,
continued supply of stars to the SBH requires some
mechanism for loss-cone repopulation.
The  most widely discussed mechanism is gravitational
encounters, which drive a diffusion in energy and
angular momentum.
The latter dominates the loss rate, since even a small
percentage change in angular momentum can put a typical 
star into the loss cone \citep{FR-76,LS-77}.

A large fraction of the stars within the SBH's
influence radius will be deflected into $r_t$ in one
relaxation time, i.e. the loss rate is roughly
$(M_\bullet/m_\star)/T_r(r_h)$.
In a collisional nucleus with $M_\bullet\approx 10^6M_\odot$, 
this is $\sim 10^6/(10^{10}$ yr$)\approx 10^{-4}$ yr$^{-1}$.
A slightly more careful calculation for the tidal disruption
rate in a $\rho\propto r^{-2}$ nucleus gives
\begin{eqnarray}
\dot N &\approx& 7\times 10^{-4} {\rm yr}^{-1}
\left({\sigma\over 70 {\rm km\ s}^{-1}}\right)^{7/2}
\left({M_\bullet\over 10^6M_\odot}\right)^{-1} \nonumber \\
&\times& \left({m_\star\over M_\odot}\right)^{-1/3}
\left({R_\star\over R_\odot}\right)^{1/4}
\label{eq:tidal}
\end{eqnarray}
where $m_\star$ and $R_\star$ are the mass and radius
of the tidally disrupted stars \citep{wang-04}.
Assuming $\rho\sim r^{-2}$ is reasonable since this
is approximately the slope observed with the 
{\it Hubble Space Telescope} at $r\gap r_h$ in
the low-luminosity galaxies that dominate the event rate
\citep{gebhardt-96}.

A handful of X-ray flares have been detected that
are candidates for tidal disruptions
\citep{komossa-99,komossa-04,halpern-04}.
The mean disruption rate computed from this
small set of events is very uncertain but is probably 
consistent with Eq.~(\ref{eq:tidal}) \citep{donley-02}.
The nuclei in which tidal disruptions have occurred
appear to remain luminous for $1-10$ yr after disruption
and possibly longer; in three of the observed events,
the luminosity decay approximately obeys the
$L_x\propto t^{-5/3}$ dependence predicted if
emission is produced during the fallback of
stellar debris onto an accretion disk 
\citep{phinney-89,evans-89}.

Classical loss cone theory \citep{LS-77,CK-78},
from which expressions like Eq.~(\ref{eq:tidal}) 
were derived,
was directed toward understanding
the observable consequences of massive black holes 
at the centers of globular clusters, which
are many relaxation times old.
As Fig.~\ref{fig:tr} shows, few galactic nuclei
are expected to be much older than $T_r$,
and loss cone dynamics in galactic nuclei can therefore be 
very different than in globular clusters.
For instance, in a nucleus that until recently contained a binary SBH,
orbits of stars with pericenters $r_{peri}\lap a_h$ will
have beeen depleted.
The time required for gravitational encounters to repopulate
these orbits is $\sim (a_h/r_h)T_r\approx (m_2/m_1) T_r$.
For $m_2/m_1=0.1$, this time exceeds $10^{10}$ yr
in most galaxies with $M_\bullet\gap 10^8M_\odot$ \citep{wang-05}.
Until the phase-space gap is refilled, tidal disruption
rates can be much lower than in a collisionally-relaxed
nucleus.

On the other hand, as Fig.~\ref{fig:triax} implies,
loss-cone repopulation in a non-axisymmetric (triaxial
or barred) nucleus can be much {\it more} efficient
than repopulation due to gravitational encounters in the spherical geometry, 
due to the presence of centrophilic (box or chaotic) orbits
in the triaxial geometry.
Numerical integrations of centrophilic orbits 
show that the rate at which a single star
experiences near-center passages with pericenter distances
$\le d$ is proportional to $d$ \citep{GB-85}.
In a $\rho\sim r^{-2}$ nucleus, the implied rate
of supply of stars to the event horizon of the SBH is 
$\sim f_c\sigma^5/Gc^2$ where $f_c$ is the fraction
of orbits that are centrophilic.
Even for modest values of $f_c$ ($\sim 0.1$),
this collisionless mechanism can supply stars to
the SBH at higher rates than collisional loss-cone
repopulation, particularly in galaxies with
$M_\bullet\gap 10^7 M_\odot$ in which two-body
relaxation times are very long \citep{poon-04b}.
In fact, loss rates in the triaxial geometry
can approach the so-called ``full loss cone''
feeding rates in spherical galaxies, 
which were invoked, in an early model, to
explain QSO activity \citep{hills-75,young-77}
and which have recently been revived
\citep{zhao-02,jordi-05}.

These arguments suggest that the mean rate
of stellar tidal disruptions in galactic nuclei
is poorly constrained by pure theory.
One way to observationally constrain the disruption 
rate is via the X-ray luminosity function of active galaxies
\citep{ueda-03,barger-05,hasinger-05}.
Assuming the $L_X\propto t^{-5/2}$ time dependence
discussed above for individual events, and convolving
Eq.~(\ref{eq:tidal}) with the SBH mass function,
one concludes \citep{mmh-06} that tidal disruptions can account for
the majority of X-ray selected AGN with soft 
X-ray luminosities below $\sim 10^{43}-10^{44}$ erg s$^{-1}$.

Nearer to home, it might be possible to search for ``afterglows''
of the most recent tidal disruption event at the Galactic
center, which could plausibly have occurred as little
as $\sim 10^3$ yr ago (Eq.~\ref{eq:tidal}).
Possible examples of such signatures include
X-ray flourescence of giant molecular clouds
\citep{sunyaev-98} and
changes in the surface properties of irradiated stars 
\citep{jimenez-06}.

\begin{acknowledgements}
I thank A. Graham for permission to reprint Fig.~\ref{fig:sersic}.
This work was supported by grants from the NSF, NASA,
and the Space Telescope Science Institute.
Some of the computations described here were carried out
at the Center for Advancing the Study of Cyberinfrastructure
at the  Rochester Institute of Technology.
\end{acknowledgements}

\bibliographystyle{aa}

\begin{thebibliography}{}

\bibitem[Alexander(1999)]{alexander-99} 
Alexander, T.\ 1999, 
\apj, 527, 835 
 
\bibitem[Amaro-Seoane et al.(2004)]{amaro-04} 
Amaro-Seoane, P., Freitag, M., \& Spurzem, R.\ 2004, 
\mnras, 352, 655 

\bibitem[Bahcall \& Wolf(1976)]{bw-76} 
Bahcall, J.~N., \& Wolf, R.~A.\ 1976, 
\apj, 209, 214 

\bibitem[Ballo et al.(2004)]{ballo-04} 
Ballo, L., Braito, V., Della Ceca, R., Maraschi, L., 
Tavecchio, F., \& Dadina, M.\ 2004, 
\apj, 600, 634 

\bibitem[Barger et al.(2005)]{barger-05} 
Barger, A.~J., Cowie, L.~L., Mushotzky, R.~F., Yang, Y., 
Wang, W.-H., Steffen, A.~T., \& Capak, P.\ 2005, 
\aj, 129, 578 

\bibitem[Baumgardt et al.(2004)]{baumgardt-04} 
Baumgardt, H., Makino, J., \& Ebisuzaki, T.\ 2004, 
\apj, 613, 1133

\bibitem[Begelman, Blandford \& Rees(1980)]{BBR-80}
Begelman, M. Blandford, R.~D. \& Rees, M.~J. 1980, 
Nature, 287, 307
 
\bibitem[Berczik, Merritt \& Spurzem(2005)]{bms-05}
Berczik, P., Merritt, D., \& Spurzem, R., 
ApJ, 633, 680

\bibitem[Berczik et al.(2006)]{berczik-06} 
Berczik, P., Merritt, D., Spurzem, R., \& Bischof, H.-P.\ 2006, 
ArXiv Astrophysics e-prints, 
arXiv:astro-ph/0601698 

\bibitem[Boylan-Kolchin et al.(2004)]{copycat-04} 
Boylan-Kolchin, M., Ma, C.-P., \& Quataert, E.\ 2004, 
\apjl, 613, L37 
 
\bibitem[Brown et al.(2005)]{brown-05} 
Brown, W.~R., Geller, M.~J., Kenyon, S.~J., \& Kurtz, M.~J.\ 2005, 
\apjl, 622, L33 

\bibitem[Brown et al.(2006)]{brown-06} 
Brown, W.~R., Geller, M.~J., Kenyon, S.~J., \& Kurtz, M.~J.\ 2006, 
ArXiv Astrophysics e-prints, 
arXiv:astro-ph/0601580 
 
\bibitem[Cohn \& Kulsrud(1978)]{CK-78} 
Cohn, H., \& Kulsrud, R.~M.\ 1978, \apj, 226, 1087 

\bibitem[Donley et al.(2002)]{donley-02} 
Donley, J.~L., Brandt, W.~N., Eracleous, M., \& Boller, T.\ 2002, 
\aj, 124, 1308 

\bibitem[Duncan \& Shapiro(1983)]{duncan-83} 
Duncan, M.~J., \&  Shapiro, S.~L.\ 1983, \apj, 268, 565 

\bibitem[Escala et al.(2004)]{escala-04} 
Escala, A., Larson, R.~B., Coppi, P.~S., \& Mardones, D.\ 2004, 
\apj, 607, 765 

\bibitem[Escala et al.(2005)]{escala-05} 
Escala, A., Larson, R.~B., Coppi, P.~S., \& Mardones, D.\ 2005, 
\apj, 630, 152  

\bibitem[Evans \& Kochanek(1989)]{evans-89} 
Evans, C.~R., \& Kochanek, C.~S.\ 1989, 
\apjl, 346, L13 

\bibitem[Favata et al.(2004)]{favata-04} 
Favata, M., Hughes, S.~A., \& Holz, D.~E.\ 2004, 
\apjl, 607, L5 
 
\bibitem[Ferrarese et al.(2006)]{acs-6} 
Ferrarese, L., et al.\ 2006, 
ArXiv Astrophysics e-prints, arXiv:astro-ph/0602297 

\bibitem[Ferrarese \& Ford(2005)]{ff-05} 
Ferrarese, L., \& Ford, H.\ 2005, 
Space Science Reviews, 116, 523 

\bibitem[Ferrarese \& Merritt(2000)]{FM-00}
Ferrarese, L. \& Merritt, D., 
\apj, {\bf 539}, L9

\bibitem[Frank \& Rees(1976)]{FR-76} 
Frank, J., \& Rees, M.~J.\ 1976, \mnras, 176, 633 

\bibitem[Freitag \& Benz(2002)]{freitag-02} 
Freitag, M., \& Benz, W.\ 2002, \aap, 394, 345  

\bibitem[Gebhardt et al.(1996)]{gebhardt-96} 
Gebhardt, K., et al.\ 1996, 
\aj, 112, 105 

\bibitem[Gebhardt et al.(2003)]{gebhardt-03} 
Gebhardt, K., et al.\ 2003, \apj, 583, 92 

\bibitem[Genzel et al.(2003)]{genzel-03} 
Genzel, R., et al.\ 2003, 
\apj, 594, 812 
 
\bibitem[Gerhard \& Binney(1985)]{GB-85}
Gerhard, O.~E., \& Binney, J. 1985,
MNRAS, 216, 467

\bibitem[Graham(2004)]{graham-04} 
Graham, A.~W.\ 2004, 
\apjl, 613, L33 

\bibitem[Graham et al.(2001)]{graham-01} 
Graham, A.~W., Erwin, P., Caon, N., \& Trujillo, I.\ 2001, 
ApJ, 563, L11 

\bibitem[Gualandris et al.(2005)]{gual-05} 
Gualandris, A., Zwart, S.~P., \& Sipior, M.~S.\ 2005, 
\mnras, 363, 223 

\bibitem[Haehnelt \& Kauffmann(2002)]{HK-02} 
Haehnelt, M.~G., \& Kauffmann, G.\ 2002, 
\mnras, 336, L61 

\bibitem[Halpern et al.(2004)]{halpern-04} 
Halpern, J.~P., Gezari, S., \& Komossa, S.\ 2004, \apj, 604, 572 

\bibitem[Hasinger et al.(2005)]{hasinger-05} 
Hasinger, G., Miyaji, T., \& Schmidt, M.\ 2005, 
\aap, 441, 417 

\bibitem[Hills(1975)]{hills-75} 
Hills, J.~G.\ 1975, \nat, 254, 295 
 
\bibitem[Hills(1988)]{hills-88} 
Hills, J.~G.\ 1988, 
\nat, 331, 687 
 
\bibitem[Hirsch et al.(2005)]{hirsch-05} 
Hirsch, H.~A., Heber, U., O'Toole, S.~J., \& Bresolin, F.\ 2005, 
\aap, 444, L61 
 
\bibitem[Holley-Bockelmann et al.(2005)]{holley-05} 
Holley-Bockelmann, K., Sigurdsson, S., Mihos, J.~C., Feldmeier, J.~J., 
Ciardullo, R., \& McBride, C.\ 2005, 
ArXiv Astrophysics e-prints, arXiv:astro-ph/0512344 

\bibitem[Holley-Bockelmann \& Sigurdsson(2006)]{holley-06} 
Holley-Bockelmann, K., \& Sigurdsson, S.\ 2006, ArXiv Astrophysics 
e-prints, arXiv:astro-ph/0601520 

\bibitem[Jimenez et al.(2006)]{jimenez-06} 
Jimenez, R., da Silva, J.~P., Oh, S.~P., Jorgensen, U.~G., 
\& Merritt, D.\ 2006, 
ArXiv Astrophysics e-prints, arXiv:astro-ph/0601527 
 
\bibitem[Komossa(2003)]{sk-03} 
Komossa, S.\ 2003, 
AIP Conf.~Proc.~686: The Astrophysics of Gravitational 
Wave Sources, 686, 161 

\bibitem[Komossa et al.(2003)]{komossa-03} 
Komossa, S., Burwitz, V., Hasinger, G., Predehl, P., Kaastra, J.~S., 
\& Ikebe, Y.\ 2003, 
\apjl, 582, L15 

\bibitem[Komossa \& Greiner(1999)]{komossa-99} 
Komossa, S., \& Greiner, J.\ 1999, \aap, 349, L45 

\bibitem[Komossa et al.(2004)]{komossa-04} 
Komossa, S., Halpern, J., Schartel, N., Hasinger, G., Santos-Lleo, M., 
\& Predehl, P.\ 2004, \apjl, 603, L17 
 
\bibitem[Kormendy \& Richstone(1995)]{KR-95} 
Kormendy, J., \& Richstone, D.\ 1995, \araa, 33, 581 

\bibitem[Lightman \& Shapiro(1977)]{LS-77} 
Lightman, A.~P., \& Shapiro, S.~L.\ 1977, 
\apj, 211, 244 

\bibitem[Makino et al.(2003)]{namura-03} 
Makino, J., Fukushige, T., Koga, M., \& Namura, K.\ 2003, 
\pasj, 55, 1163 
 
\bibitem[Makino \& Funato(2004)]{makino-04} 
Makino, J., \& Funato, Y.\ 2004, 
\apj, 602, 93 

\bibitem[Marchant \& Shapiro(1980)]{marchant-80} 
Marchant, A.~B., \& Shapiro, S.~L.\ 1980, \apj, 239, 685 

\bibitem[Marconi \& Hunt(2003)]{MH-03}
Marconi, A. \& Hunt, L.K. 2003,
AJ, 589, L21

\bibitem[Merritt(2006)]{merritt-06}
Merritt, D. 2006,
Reports on Progress in Physics,
in press

\bibitem[Merritt, Mikkola \& Szell(2006)]{MMS-06}
Merritt, D., Mikkola, S. \& Szell, A. 2006,
in preparation

\bibitem[Merritt et al.(2004)]{merritt-04} 
Merritt, D., Milosavljevi{\'c}, M., Favata, M., Hughes, S.~A., 
\& Holz, D.~E.\ 2004, 
\apjl, 607, L9 

\bibitem[Merritt \& Poon(2004)]{poon-04b}
Merritt, D., \& Poon, M.~Y. 2004,
ApJ, 606, 788

\bibitem[Merritt \& Szell(2005)]{MS-05} 
Merritt, D., \& Szell, A.\ 2005, 
ArXiv Astrophysics e-prints, arXiv:astro-ph/0510498 

\bibitem[Merritt \& Wang(2005)]{wang-05} 
Merritt, D., \& Wang, J.\ 2005, \apjl, 621, L101 
 
\bibitem[Meszaros \& Silk(1977)]{silk-77} 
Meszaros, P., \& Silk, J.\ 1977, \aap, 55, 289 

\bibitem[Mikkola \& Valtonen(1990)]{MV-90} 
Mikkola, S., \& Valtonen, M.~J.\ 1990, 
\apj, 348, 412 

\bibitem[Mikkola \& Valtonen(1992)]{MV-92} 
Mikkola, S., \& Valtonen, M.~J.\ 1992, 
\mnras, 259, 115 

\bibitem[Milosavljevi{\'c} \& Merritt(2001)]{MM-01} 
Milosavljevi{\'c}, M., \& Merritt, D.\ 2001, 
\apj, 563, 34 
 
\bibitem[Milosavljevic \& Merritt(2003)]{MM-03}
Milosavljevi{\'c}, M., \& Merritt, D. 2003,
ApJ, 596, 860

\bibitem[Milosavljevic, Merritt \& Ho(2006)]{mmh-06} 
Milosavljevic, M., Merritt, D., \& Ho, L..\ 2006, 
ArXiv Astrophysics e-prints, 
arXiv:astro-ph/0602289

\bibitem[Milosavljevi{\'c} et al.(2002)]{mm-02} 
Milosavljevi{\'c}, M., Merritt, D., Rest, A., \& van den Bosch, F.~C.\ 2002, 
\mnras, 331, L51 

\bibitem[Miralda-Escud{\'e} \& Kollmeier(2005)]{jordi-05} 
Miralda-Escud{\'e}, J., \& Kollmeier, J.~A.\ 2005, 
\apj, 619, 30 

\bibitem[Mortlock et al.(1999)]{mortlock-99} 
Mortlock, D.~J., Webster, R.~L., \& Francis, P.~J.\ 1999, 
\mnras, 309, 836 

\bibitem[Phinney(1989)]{phinney-89} 
Phinney, E.~S.\ 1989, 
IAU Symp.~136: The Center of the Galaxy, 136, 543 

\bibitem[Poon \& Merritt(2004)]{poon-04a} 
Poon, M.~Y., \& Merritt, D.\ 2004, \apj, 606, 774 

\bibitem[Preto et al.(2004)]{preto-04} 
Preto, M., Merritt, D., \& Spurzem, R.\ 2004, 
\apjl, 613, L109 
 
\bibitem[Quinlan(1996)]{quinlan-96} 
Quinlan, G.~D.\ 1996, 
New Astronomy, 1, 35 
 
\bibitem[Rees(1990)]{rees-90} 
Rees, M.~J.\ 1990, Science, 247, 817 
 
\bibitem[Salpeter(1964)]{salpeter-64} 
Salpeter, E.~E.\ 1964, 
\apj, 140, 796 

\bibitem[Sunyaev \& Churazov(1998)]{sunyaev-98} 
Sunyaev, R., \& Churazov, E.\ 1998, 
\mnras, 297, 1279 

\bibitem[Ueda et al.(2003)]{ueda-03} 
Ueda, Y., Akiyama, M., Ohta, K., \& Miyaji, T.\ 2003, 
\apj, 598, 886 

\bibitem[Uttley(2005)]{uttley-05} 
Uttley, P.\ 2005, 
ArXiv Astrophysics e-prints, arXiv:astro-ph/0508060 
 
\bibitem[Valluri et al.(2004)]{valluri-04} 
Valluri, M., Merritt, D., \& Emsellem, E.\ 2004, \apj, 602, 66 

\bibitem[Valluri et al.(2005)]{valluri-05} 
Valluri, M., Ferrarese, L., Merritt, D., \& Joseph, C.~L.\ 2005, 
\apj, 628, 137 

\bibitem[Volonteri, Haardt \& Madau(2003)]{VHM-03}
Volonteri, M., Haardt, F., \& Madau, P.,
ApJ, 582, 559

\bibitem[Wang \& Merritt(2004)]{wang-04} 
Wang, J., \& Merritt, D.\ 2004, \apj, 600, 149 

\bibitem[Young et al.(1977)]{young-77} 
Young, P.~J., Shields, G.~A., \& Wheeler, J.~C.\ 1977, 
\apj, 212, 367 

\bibitem[Yu(2002)]{yu-02}
Yu, Q. 2002,
{\it Mon. Not. R. Astron. Soc.}, {\bf 331}, 935

\bibitem[Zel'dovich \& Novikov(1964)]{novikov-64}
Zel'dovich, Y. B. \& Novikov, I. D. 1964,
Sov. Phs. Kokl., 158, 811

\bibitem[Zhao et al.(2002)]{zhao-02} 
Zhao, H., Haehnelt, M.~G., \& Rees, M.~J.\ 2002, 
New Astronomy, 7, 385 
 

\end{thebibliography}

\end{document}